\documentclass[10pt,thmsa,ukenglish]{article}
\usepackage{graphicx}

\newcommand{\figurecaption}{{\bf \figurename}\refstepcounter{figure}
{\bf \thefigure :} \small }

\begin{document}

\title{A relaxationless demonstration of the Quantum Zeno Paradox on an individual
atom}
\author{Chr. Balzer, Th. Hannemann, D. Rei\ss , Chr. Wunderlich, W. Neuhauser,
P.E.Toschek \\
{\it Institut f\"ur Laser-Physik, Luruper Chaussee 149, D-22761
Hamburg, Germany} }
\date{25.03.2002}

\maketitle

\begin{abstract}
The driven evolution of the spin of an individual atomic ion on the
ground-state hyperfine resonance is impeded by the observation of the ion in
one of the pertaining eigenstates. Detection of resonantly scattered light
identifies the ion in its upper ``bright'' state. The lower ``dark'' ion
state is free of relaxation and correlated with the detector by a null
signal. Null events represent the straightforward demonstration of the
quantum Zeno paradox. Also, high probability of survival was demonstrated
when the ion, driven by a fractionated $\pi $ pulse, was probed \textit{and
monitored} during the intermissions of the drive, such that the ion's
evolution is completely documented.
\end{abstract}

\twocolumn

\section{ Introduction}

Among the most fundamental questions posed in the continued debate
on the foundations of quantum mechanics are three issues: (1) The
nature of a quantum system in the state of temporal evolution
versus being in one of its eigenstates. (2) The nature of
quantum-mechanical measurement. (3) The emergence of the classical
world out of many interacting quantum systems. Related to all
these issues is a long-standing particular problem, namely the
temporal evolution of a quantum system under continuous or
repeated observation. From the characteristics of the unitary
rotation in phase space and von Neumann's state reduction it had
been concluded that the reiterated detection of a particular
observable of the system being involved in a quantum system's
evolution keeps projecting the system back into the state in which
it was initially prepared, and in the limit takes the evolution to
a halt [1-3]. More recently it has been argued that in order not
to confuse this effect of measurement with the effect of dynamical
intervention in the observed system, by environment or meter, this
quantum system under scrutiny must consist of an
\textit{individual} entity: On an ensemble, those two different
kinds of effects cannot be distinguished \textit{in principle}
[4-6]. Moreover, the demonstration of the strangest and most
perplexing features of
 the measurement process was said to require non-local interaction
indicated by a null result (``quantum Zeno paradox'', QZP [7]).

Recently, an experiment on the quantum evolution of an individual
atomic ion has been reported that satisfies these preconditions
[8]. The inhibition of the ion's evolutionon a weak resonance upon
repeated probing its state by reiterated attempts of making the
ion scatter resonance light was demonstrated. This evidence was
derived from the statistics of uninterrupted sequences of
\textit{equal} results, i.e., either all results in a sequence
signal ``scattered light on'', or all ``off''. The frequencies of
occurrence of such recorded sequences were determined. They agreed
with the joint probabilities for survival of the ion in its state
upon 1,2,3,... attempts of excitation or deexcitation, calculated
under the condition that the ion, after each light-driven period
of evolution, is set back, by the probing, to its initial
eigenstate. In particular, the ``off'' results that were
correlated with the ion excited into a metastable electronic state
represent a remarkable demonstration, since the corresponding
measurements satisfy the condition of quantum non-demolition
[4,8,9]. On the other hand, although the decay of that ``dark''
state via an $E2$ line is rather weak, the data nevertheless
require the consideration of relaxation with their quantitative
evaluation that is found somewhat involved.
\begin{figure}[tb]
\begin{center}
\includegraphics[scale=0.4]{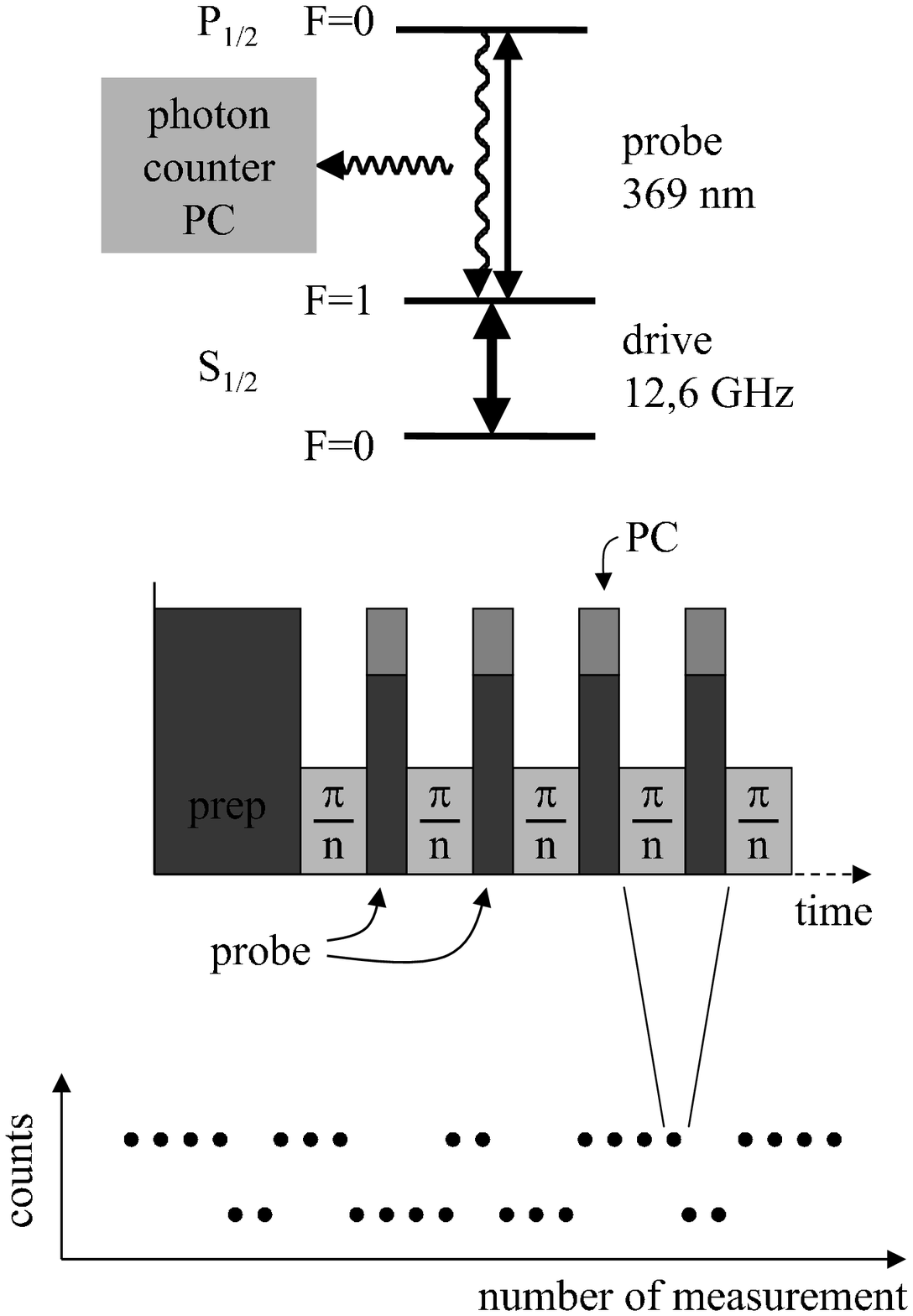}
\end{center}
\figurecaption{Relevant energy levels of $^{171}Yb^{+}$ ion,
microwave driving radiation, resonant probe light, and detection
of resonance fluorescence (top). Temporal schedule of preparation
and measurements (centre). Trajectory of results (schematic) made
up of sequences of ``on'' and ``off'' results alternating.}
\end{figure}

We report on an alternative, straightforward mi\-cro\-wave-optical
double-resonance experiment on an individual quantum system that
avoids any complication. Pulses of 12.6-GHz microwave radiation
drive the ground-state hyperfine transition of a single trapped
$^{171}$Yb$^{+}$ ion. Excitation of this ion to its $F=1$ state --
if having happened --  is monitored by laser pulses alternating
with the driving pulses: some of their light is scattered off the
ion, on the $S_{1/2}(F=1)-P_{1/2}$ resonance line. In this quantum
system, the ``dark'' state is the lower $F=0$ level of the ion's
electronic ground state and as such free of intrinsic relaxation.
Phase fluctuation of the microwave is negligible. The ``bright''
state is the upper $F=1,$ $m_F=0$ ground-state level, which is
also free of relaxation, but optical pumping into the Zeeman
sublevels $m_{F}=\pm $ $1$ by the probe light may decouple the ion
from the microwave driving and mimic energy relaxation [10]. The
relevant levels of the ion, driving and probing light, and the
laser-excited and detected resonance fluorescence are
schematically outlined in Fig 1 (top).

From recorded trajectories of the results of reiterated
measurements, the distributions of sequences of identical results
were determined. They fit in with the distributions of the
calculated joint probabilities of survival when assuming state
reduction to happen with the probing. –- An alternative strategy
involved $n$-fold irradiation of the ion by driving
$\pi/n$-pulses, intertwined by probe pulses. Recording of the
corresponding bursts of scattered probe light -– or their absence
–- allows complete characterization of the ion's evolution along
the trajectory of measurements. This strategy amounts to a
complete, or ``selective'' measurement of the ion's probability of
survival in its eigenstate. This probability increases with $n$,
and in particular the ``off'' sequences thus demonstrate the QZP.

\section{Experimental}

The experimental concept is close to a previously used one [11].
However, the source of the probe light has been modified:
Frequency-doubled light of a Ti:sapphire laser at 369 nm
wavelength with about 100 kHz bandwidth, some 10 MHz down-tuned
from resonance, was scattered on the $S_{1/2}-P_{1/2}$ ionic
resonance line. Microwave and laser parameters as well as data
acquisition were controlled in real time. The ion was occasionally
pumped into the metastable $D_{3/2}$ level, but it was immediately
repumped, into the ground state, by the 935-nm light of a diode
laser. Residual micro-motion of the ion was monitored by
phase-sensitive detection of motional fluorescence modulation, and
eliminated. Thus, the ion was localized in the node of the
trapping field with less than 10 nm deviation. Its steady-state
vibrational temperature was at the Doppler cooling $\lim $it,
below 1 mK, and well inside the Lamb-Dicke regime.

With the laser light off, the ion was irradiated by microwave pulses of
duration $\tau $ and Rabi frequency $\Omega $, such that the pulse area was $%
\theta =\Omega \tau =\pi /n,$ where $n$ is a small integer. Particular care
was taken with the precise setting of the resonance frequency and of $\theta
$. Any unwanted detuning of the microwave frequency from exact resonance was
uncovered when the ion was driven by double pulses, and afterwards probed by
a resonant laser pulse, according to the temporal version of Ramsey's
technique [11]. The results of some 500 of these measurements with stepwise
incremented temporal separation make a trajectory. With 50 samples of such
trajectories superimposed, interference fringes herald residual precession
of the ion's spin and the concomitant detuning of the microwave from
resonance, that was to be eliminated.

The pulse area was set by stepwise incrementing the length of single driving
pulses, being followed by a pulse of probe light (Fig. 2). Superimposed
trajectories of results reveal nutational oscillation of the ion's spin
whose phase is identified with the pulse area $\theta $, such that the
desired pulse length can be precisely preset, to a well-defined fraction of $%
\pi $, in contrast with the previous experiment [8].
\begin{figure}[htb]
\begin{center}
\includegraphics[scale=0.43]{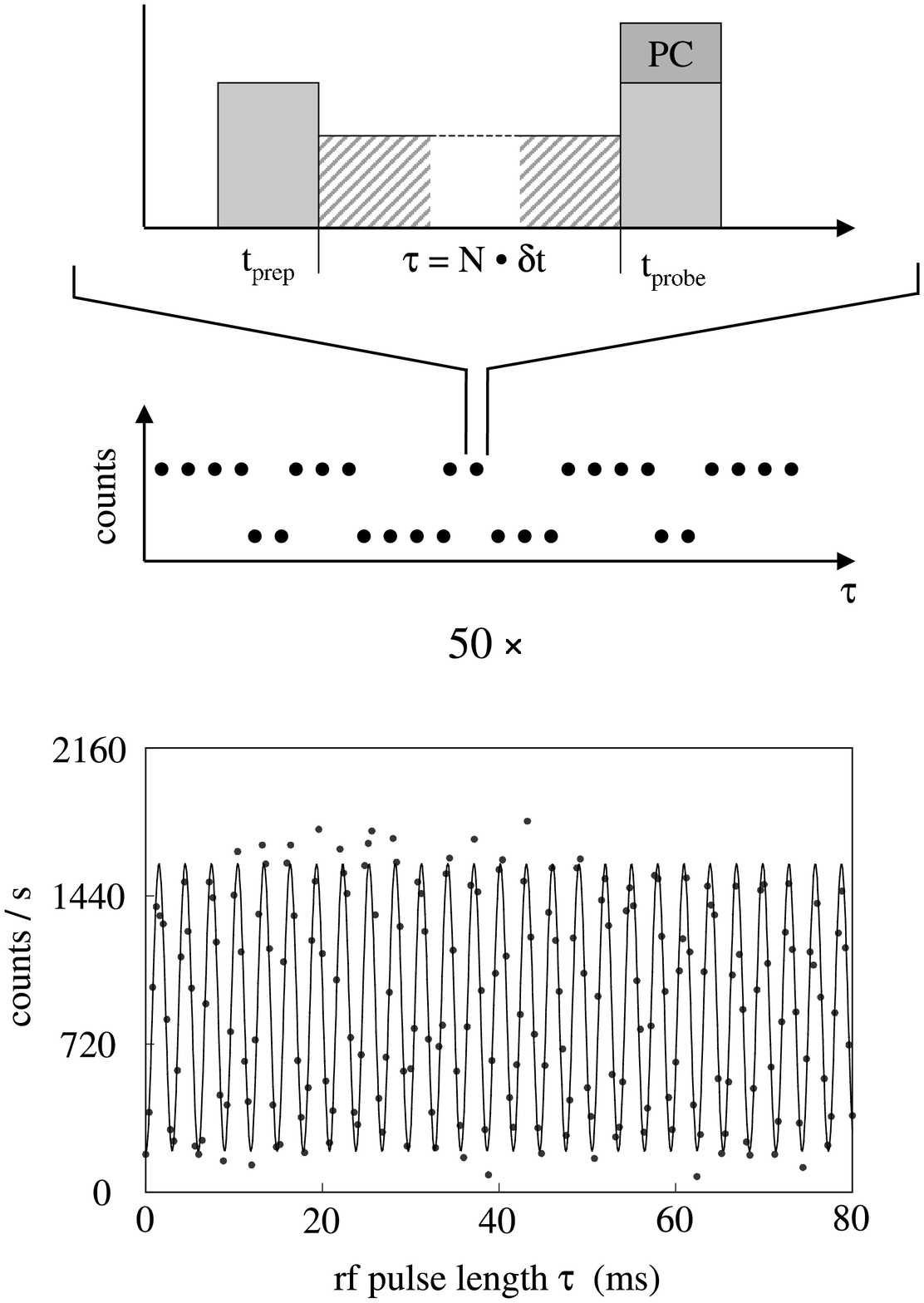}
\end{center}
\figurecaption{ Setting the pulse area of the driving radiation: A
test measurement on the single $^{171}$Yb$^{+}$ ion is made up of
a laser pulse that pumps the ion into the $F=0$ level of the
ground state, a microwave driving pulse whose duration is stepwise
increased in the next measurement, and a probe pulse with
simultaneous detection of resonant scattering. A series of 500
measurements make a trajectory; 50 superimposed trajectories show
Rabi pulsations from the ion's optical nutation. The phase of
pulsation is identified with the area $\theta =\Omega \tau $ of
the corresponding driving pulse. In a similar way, the resonant
tuning of the drive was tested in a Ramsey-type scheme, when the
driving pulse was replaced by two 2-ms-long pulses separated by
stepwise increased intermittance.}
\end{figure}

In the actual measurements, microwave driving pulses of duration
$\tau $ were separated by 1.5-ms long probe pulses of resonant
light applied to the ion that were generated when gating the cw
laser light by an acousto-optical deflector. The pulse area was
2$\pi $ when $\tau =$ 4.9 ms. The scattered light was recorded by
a photon counter gated open in synchronism with the probe pulses
(Fig. 1, centre). When a laser pulse excited a burst of resonance
scattering, the ion was considered to be on the $F=1$ hyperfine
level of its ground state; when no resonance fluorescence was
excited, the ion was supposed to be on level $F=0.$ Each pair of
driving and probing pulses represents the coherent preparation of
-- in general -- a superposition ground state, and the subsequent
read out of one of the pertaining eigenstates. Series of 10 000 of
such measurements resulted in the generation of corresponding
trajectories of results made up of alternating ``waiting
intervals'' each of which contains a quasi-random number of equal
``on'' or ``off''results (Fig. 1, bottom). The statistical
distributions of all sequences of equal results contained in these
waiting intervals was evaluated for an analysis of the retarding
effect of the measurements upon the quantum evolution of the ion
on the ground-state hyperfine resonance.

\section{Statistics of sequences of\\ equal
results}

From the data of the observed trajectories, one wants to derive a
measure for the probability of the ion's survival, during $q$
sequential attempts of driving and probing, in its eigenstate
``on'' (1) or ``off''(0). This probability is identified with the
normalized frequency of occurrence, $U(q)$, of that sequence in a
trajectory, a good approximation with a long trajectory. {\it
Matching} a value $V$ calculated under the condition of the ion's
evolution being frustrated by measurements,
\begin{equation}
U(q)/ U(1)=V(q-1),
\end{equation}
verifies QZE. The probability $V$ is easily calculated if we
assume that the driven evolution of the ion is interrupted by the
probing that leaves the ion in an {\it eigenstate}. In any
sequence of {\it equal} results, this is the same state -- namely
$|1\rangle$, correlated with ``on'', or $|0\rangle$, correlated
with ``off'' -- where the ion had been observed by the first
measurement of the sequence [8]. Since the probability of survival
after the action of one resonant driving pulse of area
$\Omega\tau$ is $p= cos^2(\Omega\tau/2)$ [12], the conditional
probability of $q$-times survival, under the above assumption, is
simply $V(q)= p^q$. Note that we need not consider, with the data
evaluation, the complications brought about by relaxation
processes, since there is no intrinsic relaxation involved with
the ground-state hyperfine transition. Thus, the probability $p$
of the ion staying in its same state as observed before, under the
action of a driving pulse, is not supposed to differ in the two
states, $p_{0}=p_{1}.$ However, although state 0 is indeed free of
relaxation, the effect of optical pumping among the Zeeman levels
of state 1 by the probe light provides the ion with some effective
energy relaxation [10].
\begin{figure}[htb]
\begin{center}
\includegraphics[scale=0.5]{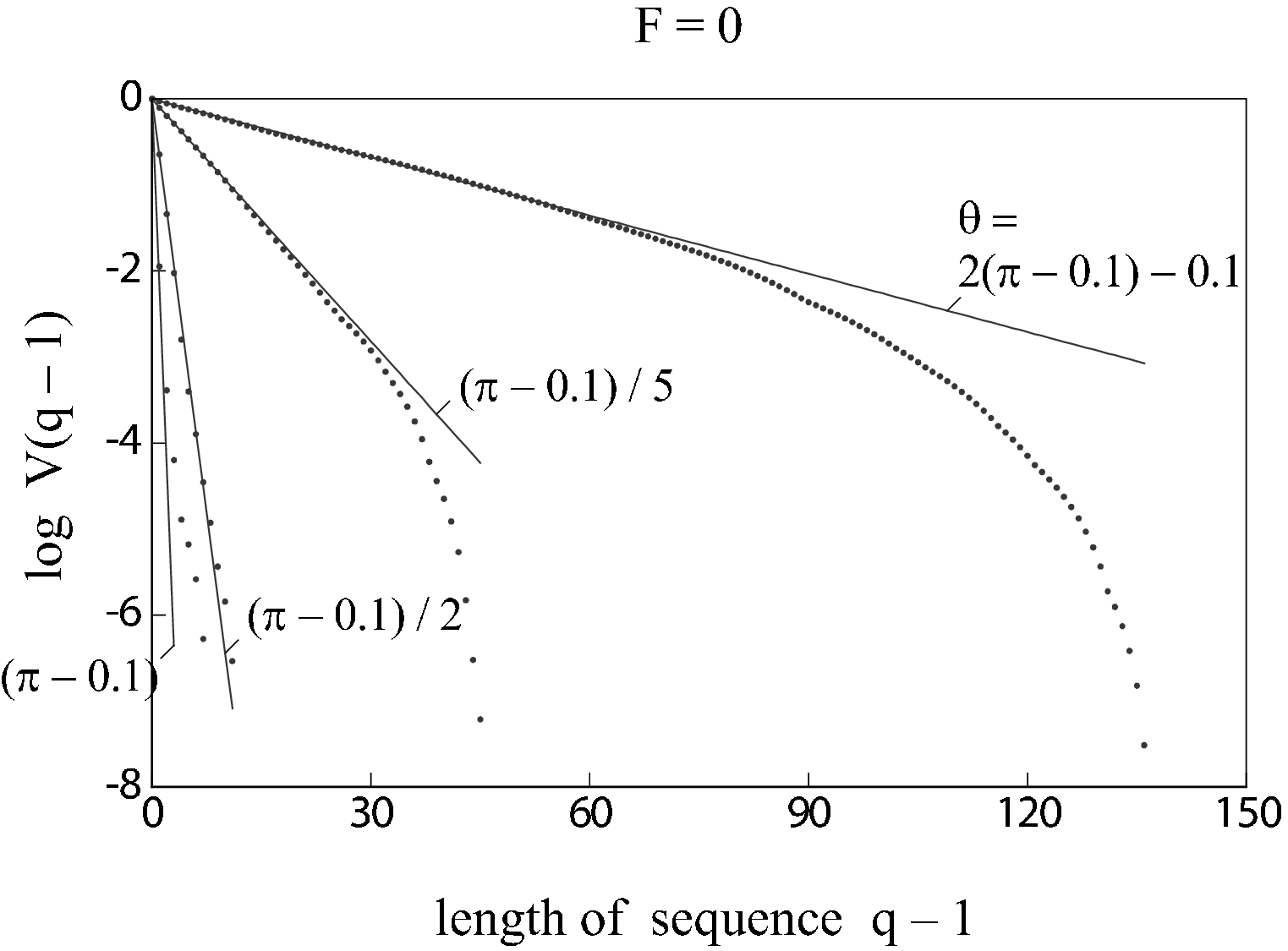}
\end{center}
\figurecaption{Probability $U(q)/U(1)$ of uninterrupted sequences
of $q$ results \textit{all} of them ``off'', when ion was
initially prepared in the ``off'' state $(F=0)$. The lines show
the distribution of probability $V(q-1) $ for the ion \textit{not}
undergoing a flip of its nuclear spin to the ``on'' state during
the entire sequence. Length of trajectories: 2000 measurements of
4.9 ms driving time and 2 ms probing time. See text.}
\end{figure}

Fig. 3 shows statistical distributions of the ``off'' sequences, namely $%
U(q)/U(1),$ on a logarithmic scale, for nominally $n=1,2,5,$ i.e., $\theta
=\pi ,$ $\pi /2,$ $\pi /5,$ and $\theta =2\pi -0.1.$ Also shown are lines
representing $V(q-1)$ that are made to fit the data by varying $\theta .$
The fitting procedure is very sensitive: deviations $\delta \theta =$ 10$%
^{-5}$ are recognizable. It turns out, that the preset areas of the
microwave $\pi $ pulses deviated from their nominal values by 3\%. At large $%
q$, the data show deficiency of long sequences, marking slightly excessive
excitation to the $F=1$ level. This feature indicates some dephasing of the
driven spin dynamics to happen on the time scale of seconds.

The distributions of the sequences of ``on'' results (not shown) also follow
straight lines on the logarithmic scale and indicate that the probability $%
p_{1}$ $=p_{0}$ does not depend on $q$. However, the variation of $U(q)/U(1)$
with $\theta $ is relatively small. These features show, even on the time
scale of only a few measurements, the effective relaxation by Zeeman pumping
and the corresponding decoupling of the ion from the driving microwave
radiation. Thus, excessive probability of long ``on'' sequences may result.

\section{Interaction with fractionated $\pi$ pulses}

We have also implemented an alternative strategy of measurement.
After a preparatory laser pulse that pumps the ion into the $F=0$
ground state, the ion was irradiated by a series of $n$ driving
pulses of area $\pi /n$. This fractionated excitation would result
in a complete $\pi $ flop of the ion into state $F=1,$ provided
that dephasing is safely negligible (Fig. 4(a) and (b)). Pulses of
probe light illuminated the ion during the $n$ $-$ $1$
intermissions between the individual driving pulses, and the
photon counter was synchronously gated open in order to register
light scattering as the signature of ion excitation after any one
of the driving pulses. The result -- scattered light on or off --
generated by the final ($n$th) probe pulse was separately
registered (Fig. 4(c)). This series of measurements, complete with
preparations and observations, was reiterated 2000/$n$ times.
\begin{figure}[tb]
\begin{center}
\includegraphics[scale=0.4]{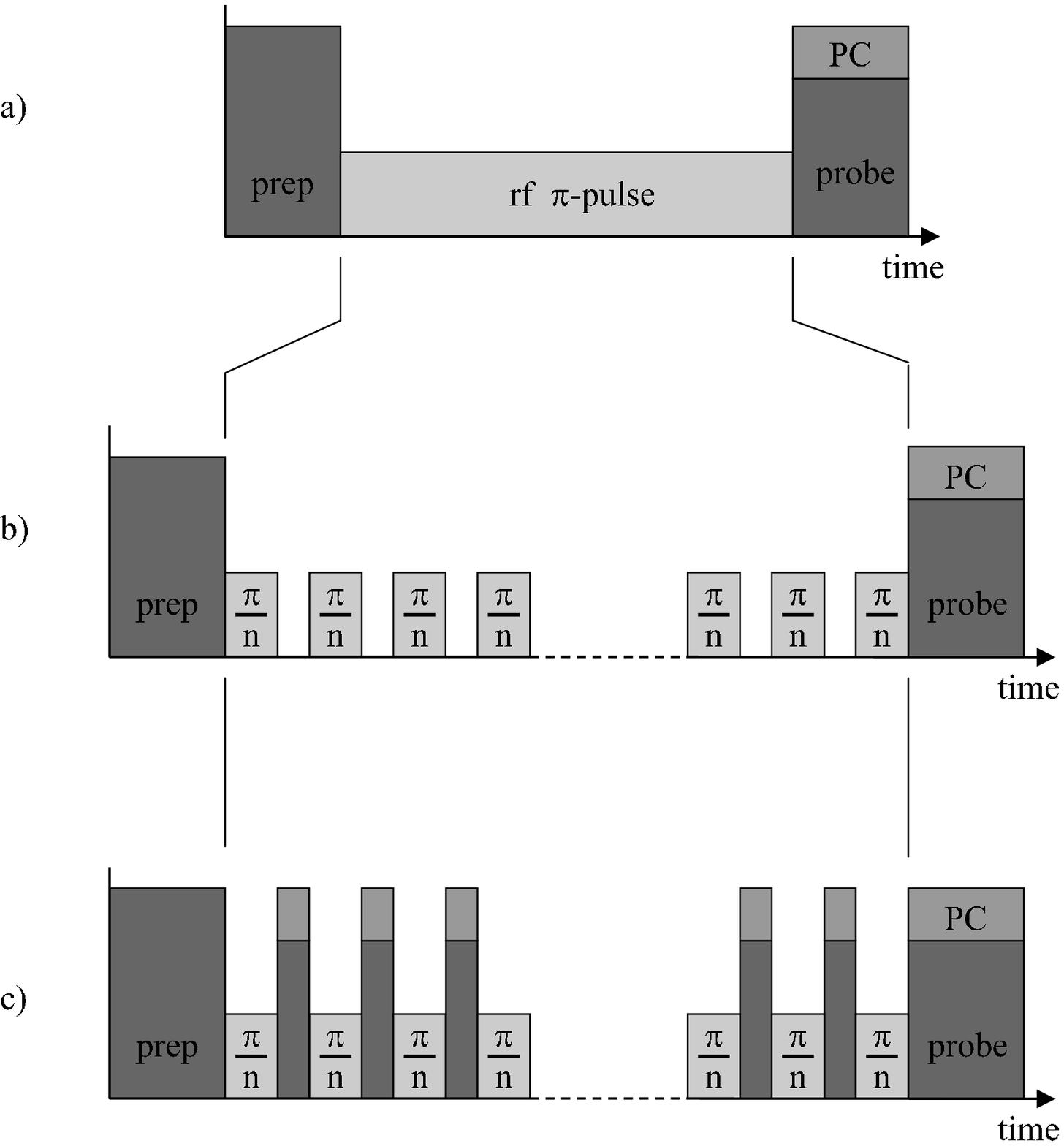}
\end{center}
\figurecaption{Scheme of measurement with one driving $\pi $ pulse
(a). Same with fractionated $\pi $ pulse ($n$-times $\pi /n)$, no
intermediate probing (b). Same, but $\pi /n$ pulses alternating
with probe pulses (c). PC: photon counting.}
\end{figure}

The recorded data may be processed in two ways:

\noindent (i) Only those series of results are considered to
represent \textit{survival} of the ion in its initial state 0,
that include no ``on'' result with any of the $n$ observations.
The number of these series, normalized by the total number of
series, approaches the probability of $n$-times survival
\begin{equation}
P_{00}(n)=\cos {}^{2n}(\pi /2n)\ .
\end{equation}
This evaluation is equivalent to discarding all series from the ensemble of
survival histories as soon as an ``on'' result shows up in any of the $n-1$
intermediate observations. The remaining subset contains series with $n-1$
times ``off'', and with both the results``off'' and ``on'' only of the final
probing, such that their sum approaches
\begin{equation}
P_{00}(n)+P_{01}(n)=cos^{2n-2}(\pi/2n)\ .
\end{equation}
The probabilities $P_{ij}(n)$ characterize ``selective''
measurements, where, $i=0$ indicates the equal outcome ``$0$'' of
the first $n-1$ measurements (out of $n$), and $j$ that of the
last one. The normalized numbers of series of $n$ measurements
that show $n$-fold survival (the frequencies of survival) are
shown as dark grey bars in the histogram of Fig. 5. The
probability of $n$-fold survival calculated from Eq. (2) is shown
as black bars. Also shown, as light grey bars, are the recorded
data with the probe pulses \textit{missing}. The corresponding
probability should vanish since effective $\pi$ pulses of the
driving microwave have been applied to the ion, which are supposed
to warrant unit transition probability, and zero probability of
survival, as long as decoherence is negligible.

\noindent (ii) All results of intermediate probing might be ignored, and the
results of the final probing normalized by the entire number of series. The
latter results would represent ``non-selective'' measurements, and the
corresponding probability of survival is
\begin{equation}
P_{00}^{(n)}=\frac{1}{2}(1+\cos ^{n}(\pi /n))\ .
\end{equation}

These values are shown by white bars underlying the black ones. A scheme of
non-selective measurements has been used in previous experimental work [13,14]. However, this
strategy is not suitable for verifying QZE [15].
\begin{figure}[htb]
\begin{center}
\includegraphics[scale=0.4]{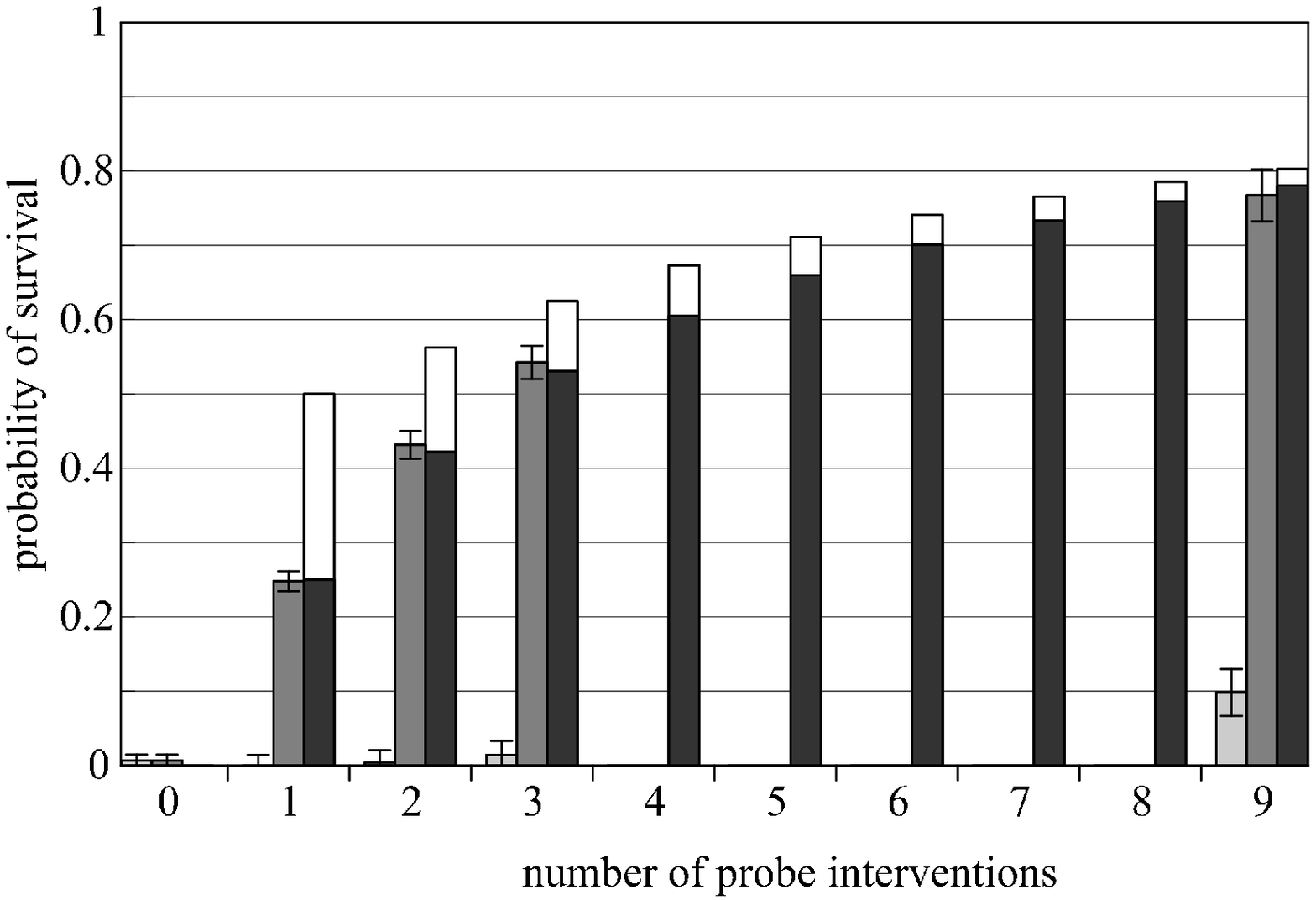}
\end{center}
\figurecaption{Probability of survival in the ``off'' state with
intermediate probing (dark grey bars), and with no probing (light
grey bars), vs. number $n$ of $\pi $-pulse partition. The
probability is evaluated by counting as ``favour\-able'' those
measurements that show only ``off'' results in each of the $n$
observations. Driving time 2.9 ms, probing time 3 ms. Probability
of survival in a ``selective'' measurement, calculated after Eq.
(2) (black bars). If evaluated from the \textit{entire} ensemble
of results, including intermediate ``on'' results, the probability
of survival would represent a non-selective measurement. This
probability is calculated after Eq. (4) (white bars).}
\end{figure}

The observed probabilities of survival are based on detected rates
of occurrence of series indentified as ``favour\-able'', i.e.
seemingly obeying the required conditions, in particular $n$
successive ``off'' detections per series. These rates have been
corrected for initial faulty ion preparation (18\% mean value) and
false detection of one of the $n$ results of photon counting in a
given series of $\pi /n$ pulses. The latter error was determined
as follows: The probability distributions of the counting rates of
``on'' (1) and ``off'' (0), vs the counted ``number of photons'',
are approximately Poissonian and overlap each other. Their
distinction is optimum when pulses that contain less than two
photon counts are identified as ``off'' results (0). In fact, the
``off'' distribution overlaps this threshold by 2\% ( mean value).
This overlap is identified as the risk of an individual false
detection, from which the rate of misinterpreting a series of $n$
measurements was determined. Multiple false detection within a
series was ignored.  The error bars of the recorded data represent
the variances of the on-off binomial distributions.

The measured and corrected frequencies of survival of the ion's
state upon the action of the $n$-times fractionated $\pi $ pulse
and intertwined probe interventions vanish at $n=0$ and increase
to 77\% at $n=9$. They agree with the values of the probability
for selective observations, calculated from Eq. (2) (black bars).
Moreover, they are at variance with probability values for
\textit{non}-selective observations, calculated according to Eq.
(4). This finding proves the QZP on an individual quantum system
without any recurrence to relaxation.

In principle, an observed set of data corresponding to the latter
probabilities $P^{(ns)}$ could be derived from the recorded data
by simply ignoring the results of $n-1$ probe interventions and
relying only to the final $n$th result. Unfortunately, these data
include intermediate ``on'' results that are affected by spurious
Zeeman pumping of the ion, by the probe light, into the sub-levels
$m_F=\pm $ $1$, outside the considered two-level system $F=0,1,$
$m_F=0$. This drawback could be avoided when optically repumping
the ion by suitably polarized light during the probe
interventions. Note that the results of such a non-selective
strategy of evaluation of the measurements are supposed to agree
with those of a previous experiment on an ensemble of particles
[13,14].

The probability of survival with lacking probe interventions vanishes,
except when driving with highly fractionated $\pi $-pulses ($n=9)$ and the
concomitant long duration of this pulse series. This 10\% survival
probability indicates the inset of some dephasing of the interaction of
microwave radiation and spin moment.

In order to better appreciate the role of decoherence, another set
of measurements was performed with the pulse length of the
intermissions between the fractionated driving pulses extended to
5 ms, but with the probe light missing. The overall duration of a
series of measurements now took substantially more time, and
dephasing -- as indicated by a finite rate of survival upon the
$\pi $-pulse excitation -- considerably increased.

\section{Conclusions}

In summary, we have performed two versions of an experiment on a
single ion determined for the demonstration of the quantum Zeno
effect with no direct interaction of quantum object with probe
radiation and meter, and by null results, i.e., of the ``quantum
Zeno paradox'' [7]. The ground-state hyperfine transition
$F=0\rightarrow 1$ of an individual $^{171}Yb^{+}$ion was
alternatingly driven by resonant microwave pulses of preset pulse
area, and probed by laser pulses that did or did not excite
resonance scattering recorded by a photon counter. Only those
series of driving and probing were counted as ``favour\-able'' for
the evaluation of the probability of survival that did
\textit{not} show excitation of resonance scattering with the
probing.

This experiment is distinguished from a previous one [8] which
required the weak spontaneous decay of the ``off'' state to be
modelled, by the complete absence of relaxation with the sequences
of results that show no scattering (``off''). These measurements
on the ion initially in the $F=0$ state are free of reaction from
the meter (``quantum non-demolition'', QND [4]). Moreover, they
are also free of relaxation. When the ion is in its $F=1$ state,
there is no intrinsic relaxation either, but appreciable
decoherence is generated when optically pumping the ion, by the
probe light, into Zeeman states unaffected by the driving
radiation.

The second version of the experiment resembles Cook's suggestion
[13]; it includes, however, two important improvements: the use of
an \textit{individual} atomic ion as the quantum system, and the
laser-excited scattering being monitored during \textit{each} of
the probing pulses. Thus, the micro-state of the ensemble of
\textit{observations} becomes completely documented, in contrast
with measurements on an ensemble of particles. This feature allows
one to generate a \textit{selective} measurement on the individual
quantum object, as is required for an unequivocal demonstration of
QZP: Only these characteristics allow one to identify
back-and-forth transitions of the quantum object during a series
of driving $\pi /n$ pulses applied to it, as well as the absence
of anti-correlated transitions in two or more individual quantum
systems of an ensemble [15]. Consequently, one is able to
distinguish the effect of the measurements, i.e., of the entangled
quantum system and probe, from that of physical intervention of
the probe upon the quantum object [7].

\smallskip

\noindent\textbf{Acknowledgements}

\medskip

This work was supported by the Hamburgische Wissenschaftliche Stiftung, and
by the Deutsche Forschungsgemeinschaft.

\end{document}